\documentclass[12pt,preprint2]{aastex61}

\usepackage{graphicx}

\usepackage{natbib}
\bibpunct{(}{)}{;}{a}{}{,} 
\usepackage{hyperref}
\hypersetup{linkcolor=red,citecolor=blue,filecolor=cyan,urlcolor=blue}
\usepackage[top=1.5in, bottom=0.in,left=2.2cm,right=0.8cm]{geometry}
\newcommand{\etal}{et al.}

\def\lvc{LIGO--Virgo }
\def\agile{{\it AGILE} }
\def\agilep{{\it AGILE}}

\def\fermigbm{{\it Fermi}-GBM }
\def\fermilat{{\it Fermi}-LAT }
\def\int{{\it INTEGRAL} }
\def\intp{{\it INTEGRAL}}
\def\intspiacs{{\it INTEGRAL}/SPI-ACS }

\def\intspi{{\it INTEGRAL}/SPI }
\def \swift{{\it Swift} }
\def \calet{{\it CALET} }
\def \kwind{{\it Konus-Wind} }
\def \inhxmt{{\it Insight-HXMT} }
\def \asat{{\it AstroSat} }
\def \maxi{{\it MAXI} }
\def \nust{{\it NuSTAR} }
\def \cha{{\it Chandra} }

\def \degmark{^\circ}

\def \arcsec {\hbox{$^{\prime\prime}$}}

\def \be {\begin{equation}}
\def \en {\end{equation}}

\def \mm { }

\def\mgw {GW170104 }

\def\gw {GW170817 }
\def\gwp {GW170817}
\def\ss {SSS17a}
\def\iau {AT~2017gfo }
\def\iaup {AT~2017gfo}

\def\t0 {$T_0$ }     
\def\tpp {$T_0$}

\def \fv {}
\def \fvv { }

\def \mt { }

\received{October 13, 2017}
\revised{October 25, 2017}
\accepted{October 25, 2017}

\shorttitle{\agile search for e.m. counterpart of \gw}

\shortauthors{F.Verrecchia \etal}

\begin{document}

\title{\agile Observations of the Gravitational-wave Source \gwp:
Constraining Gamma-Ray Emission from a NS--NS Coalescence}
%
\author{ F.~Verrecchia\altaffilmark{1,2}, M.~Tavani\altaffilmark{3,4,5},
I.~Donnarumma\altaffilmark{6,3},
A.~Bulgarelli\altaffilmark{7}, Y.~Evangelista\altaffilmark{3}, L.~Pacciani\altaffilmark{3}, A.~Ursi\altaffilmark{3},
G.~Piano\altaffilmark{3}, M.~Pilia\altaffilmark{8}, M.~Cardillo\altaffilmark{3}, N.~Parmiggiani\altaffilmark{7},
A.~Giuliani\altaffilmark{9},
C.~Pittori\altaffilmark{1,2},
F.~Longo\altaffilmark{10}, F.~Lucarelli\altaffilmark{1,2},
G.~Minervini\altaffilmark{3},
M.~Feroci\altaffilmark{3}, A.~Argan\altaffilmark{3},
F.~Fuschino\altaffilmark{7}, C.~Labanti\altaffilmark{7}, M.~Marisaldi\altaffilmark{16},
V.~Fioretti\altaffilmark{7},
A.~Trois\altaffilmark{8}, E.~Del~Monte\altaffilmark{3},
L.A.~Antonelli\altaffilmark{1},
G.~Barbiellini\altaffilmark{10},
P.~Caraveo\altaffilmark{9},
P.W.~Cattaneo\altaffilmark{11},
S.~Colafrancesco\altaffilmark{15},
E.~Costa\altaffilmark{3,6},
F. D'Amico\altaffilmark{6},
A.~Ferrari\altaffilmark{12}, P. Giommi\altaffilmark{6}, A. Morselli\altaffilmark{13},
F. Paoletti\altaffilmark{17,3}, A.~Pellizzoni\altaffilmark{8},
P.~Picozza\altaffilmark{13}, A.~Rappoldi\altaffilmark{11}, P. Soffitta\altaffilmark{3}, S.~Vercellone\altaffilmark{14},
L. Baroncelli\altaffilmark{7}, G. Zollino\altaffilmark{7}.
}
%
%

\affil {{\scriptsize $^{1}$Space Science Data Center/ASI (SSDC), via del Politecnico, I-00133 Roma, Italy} \\
 {{\scriptsize $^{2}$INAF-OAR, via Frascati 33, I-00078 Monte Porzio Catone (Roma), Italy}}\\
 {{\scriptsize $^{3}$INAF-IAPS, via del Fosso del Cavaliere 100, I-00133 Roma, Italy}}\\
 {{\scriptsize $^{4}$Dipartimento di Fisica, Univ.di Roma ``Tor Vergata'', via della Ricerca Scientifica 1, I-00133 Roma, Italy}}\\
 {{\scriptsize $^{5}$Gran Sasso Science Institute, viale Francesco Crispi 7, I-67100 L'Aquila, Italy}}\\
 {{\scriptsize $^{6}$ASI, via del Politecnico snc, I-00133 Roma, Italy}}\\
 {{\scriptsize $^{7}$INAF-IASF-Bologna, via Gobetti 101, I-40129 Bologna, Italy}}\\
 {{\scriptsize $^{8}$INAF, Osservatorio Astronomico di Cagliari, via della Scienza 5, I-09047 Selargius (CA), Italy}}\\
 {{\scriptsize $^{9}$INAF-IASF Milano, via E.Bassini 15, I-20133 Milano, Italy}}\\
 {{\scriptsize $^{10}$Dipartimento di Fisica, Universit\`a di Trieste and INFN, via Valerio 2, I-34127 Trieste, Italy}}\\
 {{\scriptsize $^{11}$INFN-Pavia, via Bassi 6, I-27100 Pavia, Italy}}\\
 {{\scriptsize $^{12}$CIFS, c/o Physics Department, University of Turin, via P. Giuria 1, I-10125,  Torino, Italy}}\\
 {{\scriptsize $^{13}$INFN Roma Tor Vergata, via della Ricerca Scientifica 1, I-00133 Roma, Italy}}\\
 {{\scriptsize $^{14}$INAF, Osservatorio Astronomico di Brera, via Emilio Bianchi 46, I-23807 Merate (LC), Italy}}\\
 {{\scriptsize $^{15}$University of Witwatersrand, Johannesburg, South Africa}}\\
 {{\scriptsize $^{16}$Birkeland Centre for Space Science, Department of Physics and Technology, University of Bergen, Norway}}\\
 {{\scriptsize $^{17}$East Windsor RSD, 25A Leshin Lane, Hightstown, NJ 08520 (USA) }}\\
}
 \email{francesco.verrecchia@ssdc.asi.it}

\begin{abstract}

The \lvc Collaboration (LVC) detected, on 2017 August 17, an
exceptional gravitational-wave (GW) event temporally consistent
within $\sim 1.7 \, s$ with the GRB 1708117A observed by \fermigbm
and \intp.
The event turns out to be compatible with a neutron star--neutron
star (NS--NS) coalescence that subsequently  produced a
radio/optical/X-ray transient detected at later times.
We report the main results of the observations by the \agile
satellite of the \gw localization region (LR) and its
electromagnetic (e.m.) counterpart.
At the LVC detection time $T_0$, the \gw LR was occulted by the
Earth.
The \agile instrument
collected useful data before and after the GW--GRB event because in
its spinning observation mode it can scan a given source many
times per hour.
The earliest exposure of the \gw LR by the gamma-ray imaging
detector (GRID) started about 935 s after \tpp.
No significant X-ray or gamma-ray emission was
 detected from the LR that was repeatedly exposed over timescales of
minutes, hours, and days before and after \gwp{\fv , also
considering Mini-calorimeter and Super-\agile data}.
%
Our measurements are among the earliest ones obtained by space
satellites on \gw and provide 
 useful constraints on the
precursor and delayed emission properties of the NS--NS
coalescence event. We can exclude with high confidence the
existence of an X-ray/gamma-ray emitting magnetar-like object with
a large magnetic field of $10^{15} \, \rm G$. Our data are
particularly significant during the early stage of evolution of
the e.m. remnant.

\end{abstract}


    \section{Introduction}

Automatic processing of \fermigbm data revealed, on 2017 August 17,
a low-fluence short gamma-ray burst (GRB; now named GRB 170817A)
detected in the 10--1000 keV range
\cite[][]{2017GCN..21506...1C,2017ApJGoldsub} {\fvv with a public notice emitted seconds after the event trigger \cite[][]{2017GCN..21520...1V}}.
%
%
\begin{figure*}[t]
   \centerline{\includegraphics[width=13.7cm, angle = 0]{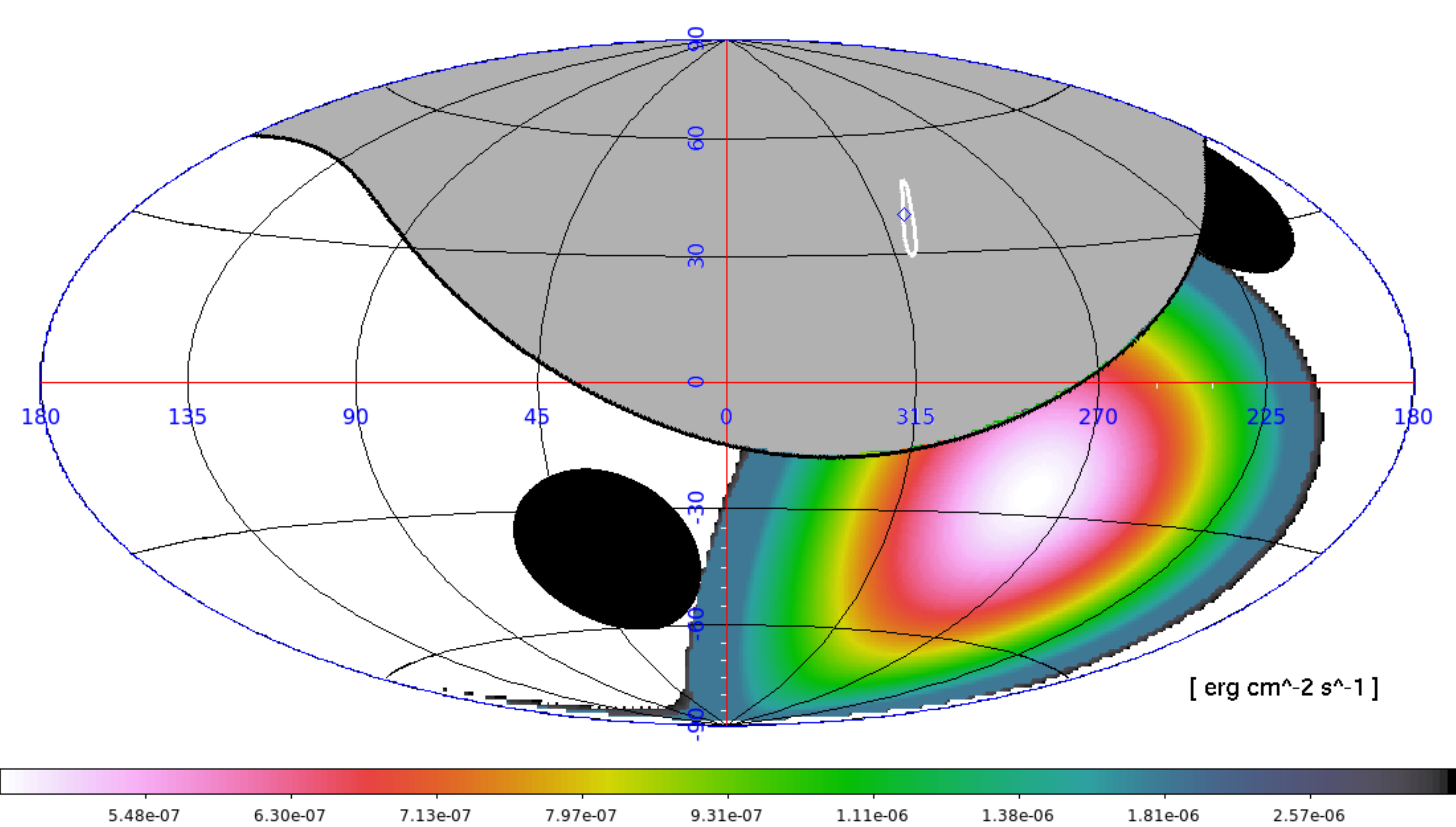}}
   \caption{\agilep-GRID $E > 30$ MeV {\fv UL} map (in $\rm erg\, cm^{-2} s^{-1}$ 
     and Galactic coordinates) based on the gamma-ray 4 s exposure at the detection time $T_0$ of \gwp.
The shadowed areas show the Earth-occulted region and the sky fraction not directly accessible by
\agile for solar panel constraints. The white contour shows the ``preliminary LAL-Inference'' 90\% c.l. LR of \gw \cite[][]{2017GCN..21527...1G}
and the blue diamond marker is at the OT \iau position.
The \agile instrument does not have significant exposure of the LR at
$T_0$, which is covered by the Earth contour.
}
 \label{fig-1}
 \end{figure*}

The \lvc Collaboration (LVC) identified a very significant
gravitational-wave (GW) event preceding the GBM trigger by {\fvv 1.7}
s. Rapid communication of this double event was promptly
issued
\cite[][]{2017GCN..21505...1E,2017GCN..21509...1E,2017GCN..21510...1S,2017GCN..21513...1S,2017GCN..21527...1G}.
GRB 170817A was  {\fv soon} confirmed by \intspiacs
(\citealp{2017GCN..21507...1S}; V. Savchenko et al., in preparation; \citealp{2017ApJLVC-GBM-INT}). This
event (now named \gwp) is very relevant for several reasons. It is
the first GW source detected in close temporal coincidence with a
short GRB, suggesting a {\fv physical} relation between the two
events. It is the first {\fv detected} gravitational-wave event
associated with a coalescence of two neutron stars. Furthermore,
it is the second GW event for which Advanced Virgo
\cite[][]{2017PhRvD..024001A} could provide crucial constraints on
the localization region restricting it to a few tens of square
degrees. These remarkable facts allowed the scientific community, for the first time, to
search for counterparts in an effective way \cite*[][ hereafter
MMA17]{2017PhRvLMMAsub}.
%
%

The {\fvv GW} event 
occurred at time
$T_0$\,=\,12:41:$0{\fvv 4.446}$ UTC
\cite[][hereafter A17b]{2017PhRvLaccep}. The event was identified
by LVC as a compact binary coalescence (CBC) of two neutron stars
(NSs) of total mass 2.82$^{+0.47}_{-0.09} \, M_{\odot}$ and 
individual masses in the range $0.86\,$--$2.26\,~M_{\odot}$. The
estimated redshift of the system is
$z\,=\,0.008^{+0.002}_{-0.003}$ corresponding to a distance {\fv of}
$40\,\pm\,{8}\, \rm Mpc$ (A17b).  
\gw is the first NS star coalescence candidate event with a
``false-alarm rate'' (FAR) less than one in 8.0$\,\times\,10^4$
years as determined by a refined analysis (A17b). 
This event is the fifth of a set of confirmed GW events detected by
LVC, GW150914 and GW151226
\cite[][]{2016PhRvD..93l2003A,2016PhRvL.116m1103A,2016PhRvL.116x1103A,2016PhRvL.116f1102A,2016PhRvD..93l2004A,2016PhRvL.116x1102A},
\mgw 
\cite[][]{2017PhRvL170104}, and the more recent GW170814
\cite[][]{2017PhRvL170814}, the first revealed using the Virgo
detector data.

 A first sky map of \gw was distributed through a LVC-GCN on 2017 August 17
  \cite[][]{2017GCN..21509...1E},
including an initial localization generated by the BAYESTAR pipeline
\cite[][]{2016PhRvD..93b4013S} and based on data from the LIGO Hanford data only.
An updated sky map was distributed about 10 hr later
 \cite[][]{2017GCN..21527...1G},
obtained using LAL-Inference \cite[][]{2015PhRvD..91d2003V}, based
on data from the three detectors LIGO Hanford, LIGO Livingston, and
Virgo. The source could be located in a sky region of 28\,$\rm deg^2$ (90\% c.l.).
%

After the LVC announcement, a multi-wavelength campaign
immediately started. The campaign involved the X-ray and gamma-ray
satellites first, and  radio/IR-optical/TeV ground observatories
at later times. The detection of a new optical transient (OT) was
first announced by the One-meter, Two-hemisphere (1M2H) team
discovered with the 1 m Swope telescope on August 18 01:05 UT
\cite[][]{2017ApJLSwopea,2017ApJLSwopeb,2017ApJLSwopec}, named
Swope Supernova Survey 2017a (\ss, now with the IAU designation
\iaup). The OT was also detected independently by five other 
teams, the Dark Energy Camera
\cite[][]{2017GCN..21530...1A,2017Cowper}, the Distance Less Than
40 Mpc Survey
\cite[L. Tartaglia et al., in preparation;][]{2017ApJLDLT40a,2017GCN..21531...1Y}, Las
Cumbres Observatory
\cite[][]{2017GCN..21538...1A,2017Arcavia,2017Arcavib}, the
Visible and Infrared Survey Telescope for Astronomy
\cite[][]{2017GCN..21544...1T,2017VISTAaccep}, and MASTER
\cite[][]{2017GCN..21546...1L,2017MASTERaccep}, REM-ROS2
\cite[][]{2017GCN..21546...1M,2017Pian,2017GCN..21592...1P}, \swift
UVOT/XRT \cite[][]{2017GCN..21612...1E,2017Evans}, and
Gemini-South \cite[][]{2017Gemsub,2017GCN..21552...1S}, and for
all see also MMA17. \iau is located at 10.$\arcsec$6 from the
early-type galaxy NGC 4993, at a distance of $\sim\,$40 Mpc.
A sequence of satellite high-energy observations started almost
immediately, with exposures of the \gw LR depending on satellite
position and operations. The X and $\gamma$-ray observations
included contributions by \calet \citep{2017GCN..21641...1N}, \kwind \citep{2017GCN..21746...1S}, \inhxmt \citep{2017GCN..21518...1L}, \asat CZTI \citep{2017GCN..21514...1B}, \agilep-GRID (see below), \fermilat \citep{2017GCN..21534...1K},
 \maxi (\citealp[][]{2017GCN..21555...1S}; S. Sugita et al. 2017, in preparation), Super-\agile (MMA17;
this work), \swift X-ray Telescope
\cite[][]{2017GCN..21612...1E,2017Evans}, \nust
\cite[][]{2017GCN..21626...1H}, \int JEM-X
\cite[][]{2017GCN..21672...1S}, and \cha
\cite[][]{2017GCN..21786...1F,2017GCN..21648...1M,2017GCN..21765...1T}.
An X--ray counterpart detection
at the OT position was reported after 9 days and confirmed
after 15 days by \cha
\cite[][]{2017GCN..21786...1F,2017ChandraM,2017ChandraT,2017GCN..21765...1T}.
Moreover, an important detection in the radio band has been reported by VLA \cite[][]{2017GCN..21545...1A,2017VLA}.
%
% ...

 \agile promptly reacted to the initial LVC
notification of \gwp, and started a quicklook analysis as data
became available within 1--2 hr as discussed below
\cite[][]{2017GCN..21564...1B,2017GCN..21526...1P,2017GCN..21525...1P,2017GCN..21785...1V}.
%
%

The \agile satellite that is 
in an equatorial orbit at an altitude of $\sim$ 500 km
\cite[][]{2009A&A...502..995T} is 
exposing 80\% of the
entire sky every 7 minutes in a ``spinning mode'', with an instantaneous field of view (FoV) of a radius of 70$\degmark$.
 The instrument consists of an imaging gamma-ray Silicon
Tracker (sensitive in the energy range 30 MeV\,--\,30 GeV),
Super-\agile (SA; operating in the energy range 20\,--\,60 keV), and the
Mini-calorimeter \citep[MCAL; working in the range 0.35\,--\,100
MeV;][]{2008A&A...490.1151M,2008NIMPA.588...17F,2009NIMPA.598..470L}
with an omni-directional FoV and self-triggering capability
in burst mode for various trigger timescales. The
anti-coincidence (AC) system completes the instrument \cite[for a
summary of the \agile mission features, see ][]{2009A&A...502..995T}.
The combination of Tracker, MCAL, and AC working as a gamma-ray
imager constitutes the \agilep-GRID. The instrument is capable of
detecting gamma-ray transients and GRB-like phenomena on
timescales ranging from submilliseconds to tens--hundreds of
seconds.

\begin{figure*}[ht!]
   \centerline{\includegraphics[width=13.7cm, angle = 0]{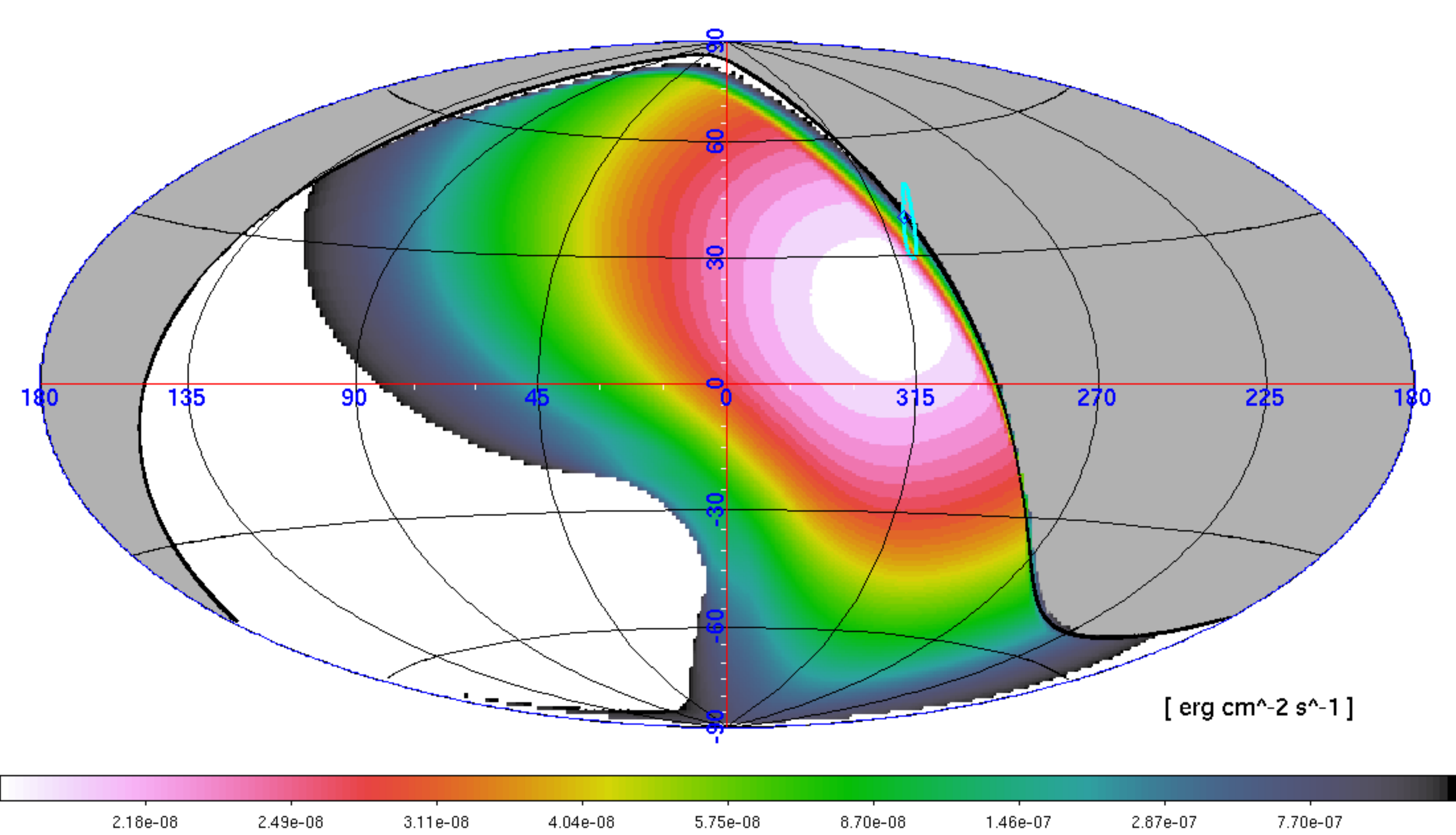}}
   \centerline{\includegraphics[width=13.7cm, angle = 0]{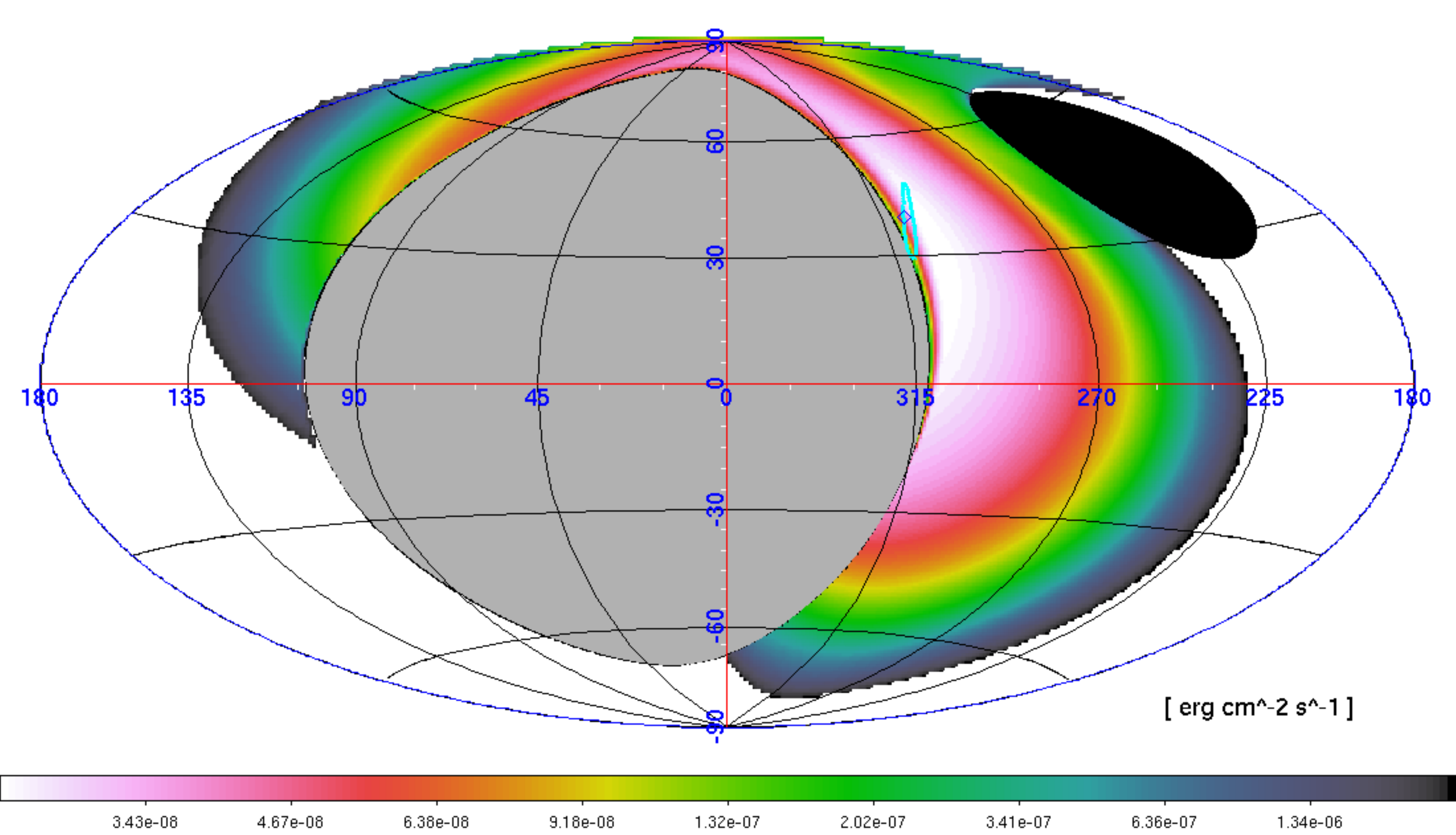}}
    \caption{
    \agilep-GRID $E > 30$ MeV {\fv UL} maps (in $\rm erg\, cm^{-2} s^{-1}$ and Galactic coordinates)
for the 150 s exposures nearest to the $T_0$, centered at --1260 and +1010 s (times refers to $T_0$). The ``preliminary LAL-Inference''
90\% c.l. LR of \gw \cite[][]{2017GCN..21527...1G} is shown in cyan color, and the blue diamond marker is at the optical transient \iau position.
}
 \label{fig-2}
\end{figure*}

The \agile instrument
has important characteristics for 
observations of large GW source LRs:
a very large FoV of the GRID (2.5 sr); 80\% of the whole sky that can be exposed every 7 minutes;
 100--150 useful passes every day for any region in the accessible sky; a gamma-ray exposure
of $\sim$\,2 minutes of any field in the accessible sky every 7
minutes; sensitivity of $\sim 10^{-8} \, \rm erg \, cm^{-2} \, s^{-1}$
above 30 MeV for typical single-pass of unocculted sky regions; a
submillisecond MCAL trigger for very fast events in the range
0.4\,--\,100 MeV; hard X-ray (18\,--\,60 keV) triggers of GRB-like
events with a localization accuracy of 2--3 arcmin in the SA FoV
($\sim 1$ sr) when operating in imaging mode.

Satellite data are currently transmitted to the ground on average
every passage over the ASI Malindi ground station in Kenya
{\fv and delivered to the \agile Data Center (ADC; part of the ASI Space Science Data Center)}.
Scientific data are then processed
 by a {\fv fast} dedicated pipeline (\agile GW, AGW),
recently enhanced for the {\mt search of electromagnetic (e.m.)
counterparts of GW sources}. \agile data processing can typically
produce an alert for a transient gamma-ray source and/or GRB-like
events within 20 minutes--2 hr from satellite onboard acquisition
depending on orbital and satellite parameters
\cite[][]{2013NuPhS.239..104P,2014ApJ...781...19B}.

In this Letter we present the main results of the analysis of \agile
data concerning \gwp. Sect.~2 presents the results on gamma-ray
emission above 30 MeV from \gwp. Sect.~3 presents the results of
Super-\agile and MCAL observations. We discuss our measurements of
\gw and their implications in Sect. 4.

\section{Gamma-Ray Observations of \gw}

\subsection{Prompt Emission}

%
\begin{figure*}  [ht!]
   \centerline{\includegraphics[width=\textwidth]{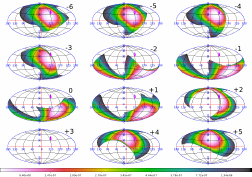}}
\caption{\agilep-GRID sequence of 12 passes over the OT \iau
obtained during the period (--2813 s, +2435 s) with respect to $T_0$. The maps report the
gamma-ray flux $3\,\sigma$ upper limits (in $\rm erg\, cm^{-2} s^{-1}$) in the energy range 30 MeV--10 GeV, with the lowest values at the OT \iau position being $ UL = 1.9 \times 10^{-8} \rm \, erg \,cm^{-2} \, s^{-1}$. The sequence shows all of the 150 s integration maps for all the single
spinning rotations  reported in Table~\ref{tab-1}. The {\fvv magenta} contour corresponds to the \gw LR according
to the ``preliminary LAL-Inference'' 90\% c.l. \cite[][]{2017GCN..21527...1G}, while the \iau position
is marked with the blue diamond symbol.
} \label{fig-3}
\end{figure*}

{\mt The \agile satellite was occulted by the Earth at the moment
of the LVC detection. Therefore, there are no prompt data obtained
by the GRID at the position of the OT.}
%
%
Figure~\ref{fig-1} shows the gamma-ray sensitivity map above 30 MeV
at the \gw trigger time, with the Earth completely covering the \gw LR (obtained from the 90\% LR
extracted from the refined localization map; A17b).
%
{\mt 
Since the \agile satellite rotates every $\sim 7$ minutes around
the axis pointed toward the Sun, the GRID obtained exposures of
the \gw LR for each satellite revolution not affected by Earth
occultation or SAA passages. We consider here three types of time
intervals}
preceding and following the $T_0$ {\mt with different integration
timescales}: (a) short timescales, with 150 s
integrations covering the interval [--3100, +2900]~s (times refer
to $T_0$); (b) medium timescales, with integrations
ranging from 1 hr to 1 day covering the interval [--1, +1] day;
(c) long timescales, with integrations ranging from 1 day
to 50 days covering the interval [--101, +21] days.

\subsection{Early Observations with Short Timescales}

{\mt For the relatively short exposures during the first passes of
the \gw LR 
"GRB detection mode", that maximizes the detection of relatively
short gamma-ray transients lasting a few tens of seconds \cite[as
already applied to the short GRB 090510;][] {2010ApJ...708L..84G}.
We show in Figure \ref{fig-3} the sequence of 150 s duration passes
over the LR within --$\,3100 \rm \,s < t < +\,2900$ s, and we
summarize in Table~\ref{tab-1} the 2\,$\sigma$ flux UL obtained at the OT
position. We label time intervals of the highest GRID exposure
with progressive numbers with respect to the prompt interval
$\Delta T_0$. The intervals nearest in time to T$_0$ are the
$\Delta T_{-3}$ and $\Delta T_{+2}$ ones.
We then obtain $2\,\sigma$ flux upper limits in the 30 MeV\,--\,10 GeV band, with
150 s integrations within the time interval [--3100, +2900]~s.

The values for the integrations nearest in time to $T_0$ are (see Figure \ref{fig-2}):
$UL_{-3} = 5.5 \times 10^{-7} \rm \, erg \, cm^{-2} \, s^{-1}$ for
the interval [--1335, --1175]~s, $UL_{+2} = 4.3 \times 10^{-8} \rm
\, erg \, cm^{-2} \, s^{-1}$ for the interval [+935, +1085]~s{\fvv , the first one being 
obtained when the OT was near the Earth limb, so that 
the effective exposure is lower than that in other intervals}.
\begin{table}[ht]
\begin{center}
{\footnotesize
{{\bf Table 1}\\ Analysis of Individual Passes over the OT
Position in the GRB Detection Mode} \vskip.2cm
\begin{tabular*}{0.48\textwidth}{lcccc}
\noalign{\hrule height 0.7pt\vskip.028cm}
\noalign{\hrule height 0.65pt\vskip.1cm}
 Interval & Start   &  $\Delta\,t$ &  2\,$\sigma$ UL  &  \multicolumn{1}{l}{Comments} \\
 Number   & Time$^{a}$ &  ~$^{b}$ & 30\,MeV--10\,GeV$^{c}$ & \\
    &  (s)  & (s)  &  ($10^{-8} \, \rm erg \, cm^{-2} \, s^{-1}$) &  \\
\noalign{\vskip.1cm\hrule height 0.7pt\vskip.1cm}
 --6& --2663 & 150 &  1.9 &  ...  \\
 --5& --2213 & 150 &  1.9 &  ...  \\
 --4& --1763 & 150 &  1.9 &  ...  \\
 --3& --1335 & 150 & 54.5 &  ...  \\
 --2& --1013 & 150 &  ... & \multicolumn{1}{l}{LVC source occul-} \\
    &        &     &  ... & \multicolumn{1}{l}{ted by the Earth}  \\
 --1& --563  & 150 &  ... & \multicolumn{1}{l}{LVC source occul-} \\
    &       &     &  ... & \multicolumn{1}{l}{ted by the Earth}  \\
   0&  0    & 150 &  ... & \multicolumn{1}{l}{LVC source occul-} \\
    &       &     &  ... & \multicolumn{1}{l}{ted by the Earth}  \\
  +1&  +335 & 150 &  ... & \multicolumn{1}{l}{LVC source occul-} \\
    &       &     &  ... & \multicolumn{1}{l}{ted by the Earth}  \\
  +2&  +935 & 150 &  4.3 &   ... \\
  +3& +1385 & 150 &  ... & excluded due to \\
    &       &     &  ... & SAA passage \\
  +4& +1835 & 150 &  2.4 &  ...  \\
  +5& +2285 & 150 &  2.2 &  ...  \\
\noalign{\hrule height 0.65pt\vskip.1cm}
 ... & --3100 & 3100 &  0.6  &  ... \\
 ... & +0    & 2900 &  0.5  &  ... \\
\noalign{\hrule height 0.7pt}
\end{tabular*}
}
\end{center}
\vskip-.2cm
\noindent {\bf \scriptsize Notes.\\}
$^{a}$~{\scriptsize Start time of the time interval with respect to $T_0$, t\,--\,$T_0$, in seconds.\\}
$^{b}$~{\scriptsize Integration time in s.\\}
$^{c}$~{\scriptsize $2\,\sigma$ flux upper limit obtained with the ``GRB detection mode'' for emission at the \iau position
with integrations of 150 s except for the last two rows obtained with exposures of 3100 and 2900 s before and after T$_0$.}

 \label{tab-1}
\end{table}
%
For a 2665 s integration including the $\Delta T_{+2}$ interval we
obtain: $UL_{2ks} = 3.9 \times 10^{-9} \rm \, erg \, cm^{-2} \,
s^{-1}$ for the overall time interval\footnote[19]{The integration
time of 2665 s corresponds to a lower effective exposure time of
$\sim 400$ s because of the satellite revolutions in spinning mode
(see Table 1).} [+935, +3600]~s. In Figure~\ref{fig-4}, we report our
\begin{figure}[ht]
\begin{center}
   \includegraphics[clip=true,width=\linewidth]{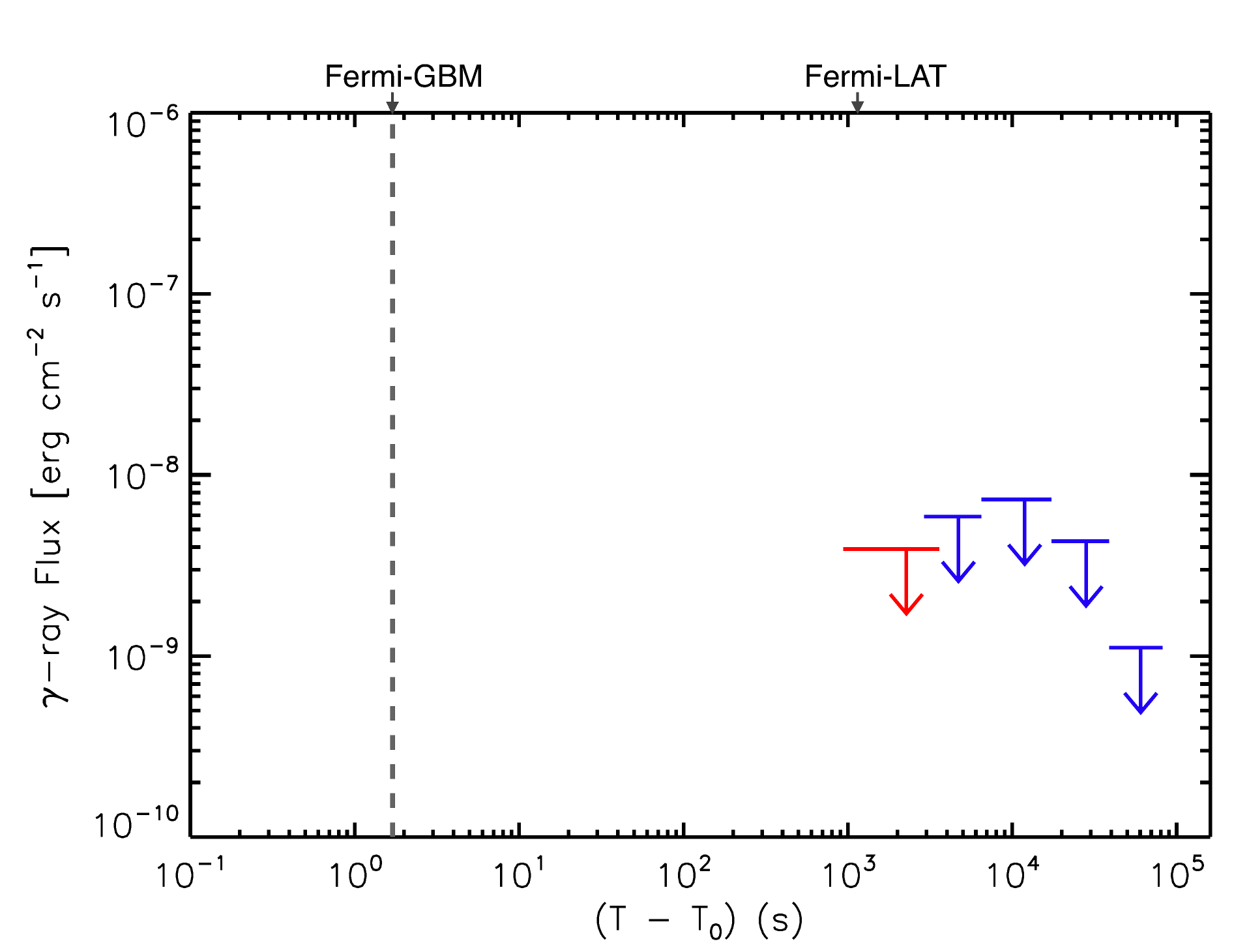}
\end{center}
 \vspace*{-0.5cm}
  \caption{We show the \agilep-GRID gamma-ray
    2\,$\sigma$ ULs obtained at the \iau position and at different times with integrations of 2665 s,
marked in red color, and integrations ranging from 1 to 12 hours, marked
in blue color. }
 \label{fig-4}
\end{figure}
gamma-ray ULs obtained at the times reported in Table 1 with the
indication of the GRB 170817A \fermigbm detection time and of the
\fermilat observation. No other gamma-ray observation of the \gw
LR is available in the range 30 MeV\,--\,30 GeV.

\subsection{Search for Precursor and Delayed Gamma-Ray Emission}

%
%
\agile was uniquely able to collect data on the \gw LR during
%
selected time intervals preceding and following the prompt event.
We carried out a search for transient gamma-ray emission on
integrations smaller than one day (see Table \ref{tab-2}) during
the hours and days immediately following or preceding the prompt
event (see Figure \ref{fig-5}).
Finally we also performed a search on longer timescales up to 100
days before and 20 days after the event. Table \ref{tab-3}
summarizes our results and ULs for these long integrations.
No significant gamma-ray emission in the \gw LR was detected over
integrations of hours up to 50 days at covering times from
T$_0$\,--\,101 days to T$_0$\,+21 days.
\begin{table}[ht]
\begin{center}
{\footnotesize
{{\bf Table 2\\} Analysis of GRID Data over Medium-timescale Integration Times at the OT Position} \vskip 0.3cm
\begin{tabular*}{0.48\textwidth}{cccccc}
\noalign{\hrule height 0.7pt\vskip.028cm}
\noalign{\hrule height 0.65pt\vskip.1cm}
 Interval & Start   &  $\Delta\,t$ &  2\,$\sigma$ UL    \\
 Number   & Time$^{a}$~ & $^{b}$~ & $^{c}$ \\

    &   (ks)   & (hr) & ($10^{-9} \, \rm erg \, cm^{-2} \, s^{-1}$)    \\
\noalign{\vskip.1cm\hrule height 0.7pt\vskip.1cm}

--10 & --32.9 & 12  &   0.66  \\
 --9 & --11.3 & 6   &   1.11  \\
 --8 & --6.7  & 3   &   1.66  \\
 --7 & --3.1  & 1   &  10.07  \\

\noalign{\hrule height 0.65pt\vskip.1cm}
  +6 & +2.9  & 1   &   5.89 \\
  +7 & +6.5  & 3   &   7.33 \\
  +8 & +17.3 & 6   &   4.31 \\
  +9 & +38.9 & 12  &   1.11 \\

\noalign{\hrule height 0.65pt\vskip.1cm}
 ... & --86.4  & 24  &  0.55   \\
 ... & +0     & 24  &  1.47  \\
\noalign{\hrule height 0.7pt}
\end{tabular*}
}
\end{center}
\vskip-.2cm
\noindent {\bf \scriptsize Notes.\\}
$^{a}$~{\scriptsize Start time of the interval with respect to $T_0$, t\,--\,$T_0$, in ks.
The second table part reports the UL on integrations on the total time covering all the pre-/post-T$_0$ intervals.\\}
$^{b}$~{\scriptsize Integration time, in hr.\\}
$^{c}$~{\scriptsize $2\,\sigma$ flux upper limit ($10^{-9} \, \rm erg \, cm^{-2} \, s^{-1}$) obtained for emission in the range 100
MeV\,--\,10 GeV with the \agile maximum likelihood analysis at the \iau position.\\}
 \label{tab-2}
\end{table}

The long-timescale GRID data analysis included two further
searches for transient gamma-ray detections with equal
integrations of 1000 s, 12 hr, 1 day, and 2 days. In particular, we
have performed a search for gamma-ray emission from 2017 August 1 
to August 31, at a 1000 s integration timescale. On 2017 
August 24, 09:31:39 UTC, using
the method described in \cite{1983Li&Ma}, we detected  a 
weak gamma-ray source {\fvv spatially} coincident with the \gw OT at
$4.2\,\sigma$ pre-trial significance.
Our analysis leads to a marginal post-trial significance of
$3.1\,\sigma$ for this source.
\begin{figure*}[ht]
   \centerline{
   \includegraphics[width=9.4cm, angle = 0]{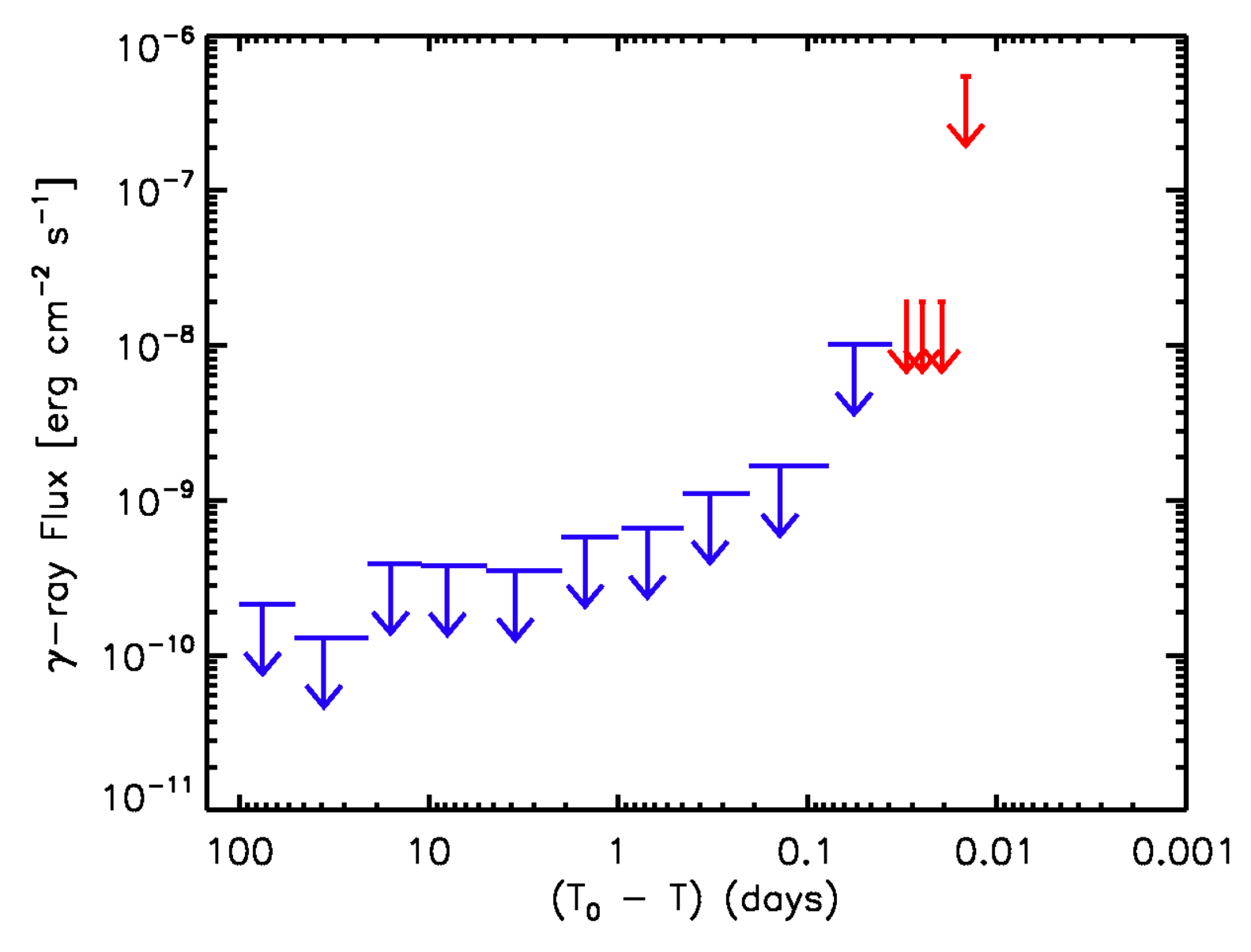}
 \hspace{-0.4cm}
   \includegraphics[width=9.2cm, angle = 0]{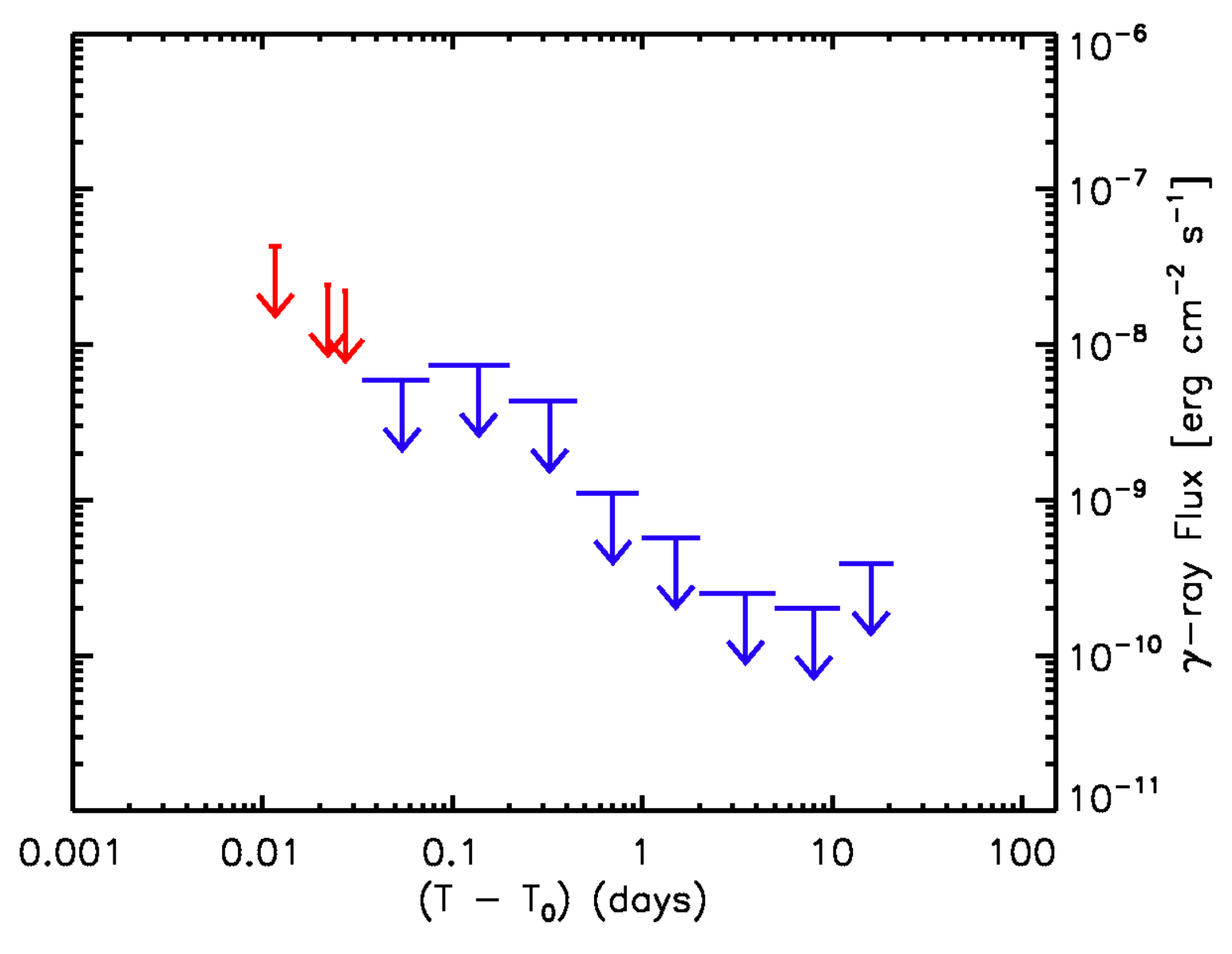}}
    \caption{\agilep-GRID gamma-ray 2\,$\sigma$ {\fvv flux} ULs obtained at the OT position
at different times and integrations. Integrations of {\fv 150} s in the 30 MeV -- 10 GeV band are marked in red {\fv color}, while those ranging from 1 hour to 100 days in 100 MeV\,--\,10 GeV are marked in blue {\fv color}.
On the left the ULs preceding $T_0$ till T$_0$--101 days, and on the right the ones following it till T$_0$+21 days.}
 \label{fig-5}
\end{figure*}
Similar searches on 1 and 2 day integration timescales over [--10,
+10] days time interval have been performed discarding low-exposure integrations, 
and executing the \agile maximum likelihood
analysis \cite[][]{1996Mattox,2012Bulgarelli} at the OT position.
Another low-significance source {\fvv spatially} coincident with the OT
is found in a 1 day integration starting at August 9, 00:00:00
UTC, with a pre-trial significance of 3.3\,$\sigma$.
%

\section{SA and MCAL observations}

The SA detector \cite[][]{2007NIMPA.581..728F} observed
the location of \gw starting at 2017 August 18 01:16:34.84 UTC, that
is, with a $\Delta$\,t $=$\,0.53 days with respect to the LVC
trigger time restarting its science observations after a 12 hr
period of "idle mode", due to telemetry-saving requirements at
mission level. We report here the analysis for the first available
orbit after the idle time interval. The location of \gw thus crossed
the FoV of the SA {\mt detector} at 23.5$\degmark$
off-axis in one coordinate (X), while scanning the entire FoV in
the orthogonal coordinate (Y). This is a rather unfavorable
viewing angle for SA, greatly reducing its effective area and
sensitivity thereof. We restricted our analysis between --15$\degmark$
and +15$\degmark$ in the Y-coordinate, exposing on average 32\% of the
peak effective area. The observation of the GW source is composed
by a set of 15 time intervals (one per satellite rotation), each
one lasting about 40 s, for a total net exposure time of 573~s. No
X-ray source was detected at the location of \gwp, with a
3\,$\sigma$ UL in the 18--60 keV energy band of $3.0\,\times
10^{-9}\, \rm erg\,cm^{-2}\,s^{-1}$.

As the OT is de-occulted and accessible by MCAL near +935 s, we
obtain flux ULs with the MCAL detector at times --1328 s and +1023
s.
No evidence of triggered or untriggered significant emission
detected by MCAL was measured.
{\mm We obtained a 2\,$\sigma$ fluence UL in the energy band 400
keV\,--\,100 MeV,
UL = $3.1\,\times 10^{-7} \rm \, erg \, cm^{-2}$ at times --1328~s
and +1023~s, }
 using a power-law model with a photon index of 1.4.

\section{Discussion and Conclusions}

\agile contributed in a significant way to the multifrequency
follow-up observations of the exceptional event \gw (MMA17).
Despite the Earth occultation of the LR at $T_0$, \agile obtained
very relevant X-ray and gamma-ray constraints on \gwp. As reported
in \citet{2017GCN..21564...1B} and MMA17,
 the earliest gamma-ray
imaging detector data were obtained near $T_1 = +935$ s. Table 1
and Figure~\ref{fig-3} show  the results of the first useful passes
over the \gw LR. Of particular relevance to our discussion is the
upper limit to gamma-ray emission above 30 MeV obtained with an
integration of about 2600 s after $T_1$, $F_{\gamma} =
3.9\,\times\,10^{-9} \rm erg\, cm^{-2}\, s^{-1}$. For a distance
$d = 40 \, \rm Mpc$ this translates to a limiting isotropic
gamma-ray luminosity $L_{iso_{\gamma}} = 7.8 \cdot 10^{44} \, \rm
erg \, s^{-1}$ {\fvv in the range 30 MeV\,--\, 10 GeV}. {\mm It is interesting to note that the peak
isotropic luminosity of GRB 170817A in the 10--1000 keV band
detected by \fermigbm is $L_{iso,GBM} \simeq 6 \cdot 10^{46} \,
\rm erg \, s^{-1}$
(\citealp[][]{2017ApJLVC-GBM-INT,2017ApJGoldsub}; MMA17).}
\begin{table}[ht]
\begin{center}
{\footnotesize
{{\bf Table 3\\}  Long-timescale Integration Times Analysis at the OT Position} \vskip 0.3cm
\begin{tabular*}{0.48\textwidth}{cccccc}
\noalign{\hrule height 0.7pt\vskip.028cm}
\noalign{\hrule height 0.65pt\vskip.1cm}
 Interval & Start   &  $\Delta\,t$ &  2\,$\sigma$ UL    \\
 Number   & Time$^{a}$~ & $^{b}$~  & $^{c}$  \\
    &   (days)  &  (days) &  ($10^{-10} \, \rm erg \, cm^{-2} \, s^{-1}$)    \\
\noalign{\vskip.1cm\hrule height 0.7pt\vskip.1cm}
 --16 & --101 & 50 &  2.12 \\
 --15 & --51  & 30 &  1.30 \\
 --14 & --21  & 10 &  3.88 \\
 --13 & --11  & 6  &  3.80 \\
 --12 & --5   & 3  &  3.52 \\
 --11 & --2   & 1  &  5.79  \\

\noalign{\hrule height 0.65pt\vskip.1cm}
  +10 & +1  & 1  &  5.71  \\
  +11 & +2  & 3  &  2.52 \\
  +12 & +5  & 6  &  2.01  \\
  +13 & +11 & 10  & 3.88 \\

\noalign{\hrule height 0.7pt}
\end{tabular*}
}
\end{center}
\vskip-.2cm
\noindent {\bf \scriptsize Notes.\\}
$^{a}$~{\scriptsize Start time of the interval with respect to $T_0$, t\,--\,$T_0$, in days.\\}
$^{b}$~{\scriptsize Integration time, in days.\\}
$^{c}$~{\scriptsize $2\,\sigma$ flux upper limit ($10^{-10} \, \rm erg \, cm^{-2} \, s^{-1}$) obtained for emission in the range 100
MeV\,--\,10 GeV with the \agile maximum likelihood analysis at the \iau position and for a mean power-law spectrum with photon index --2.}
 \label{tab-3}
\end{table}
 {\mt Figures \ref{fig-3} and \ref{fig-5} show
the overall trend of the gamma-ray upper limits obtained by the
\agilep-GRID at later times. {\mt In terms of upper limits of the
gamma-ray luminosity, the results of Figure~\ref{fig-6} indicate a
range of isotropic luminosities between $10^{43}$ and $10^{45} \,
\rm erg \, s^{-1}$}. Furthermore, the imaging SA provided
\begin{figure}[ht]
    \centerline{\includegraphics[width=9.5cm, angle = 0]{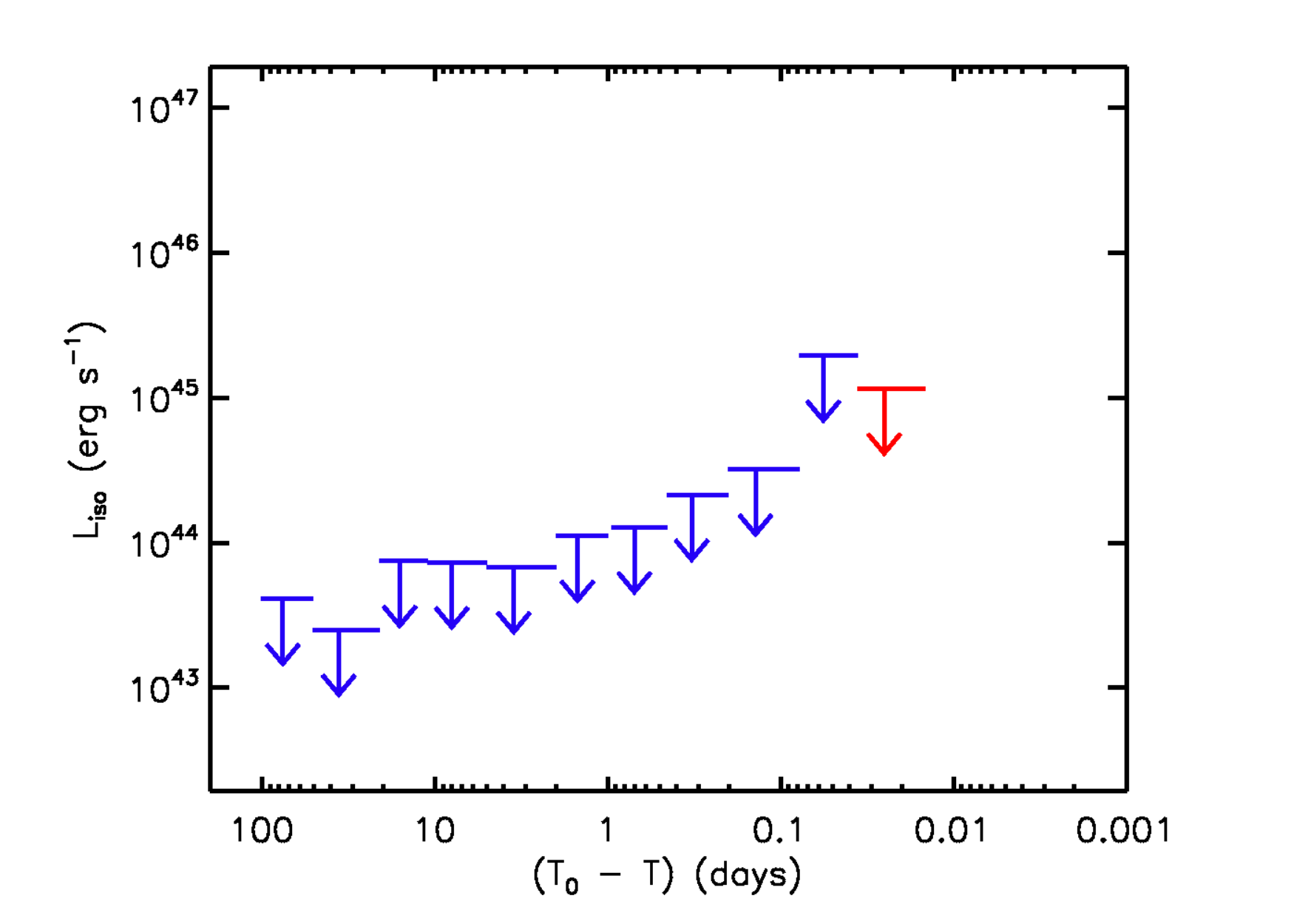}}
   \centerline{
 \includegraphics[width=9.5cm, angle = 0]{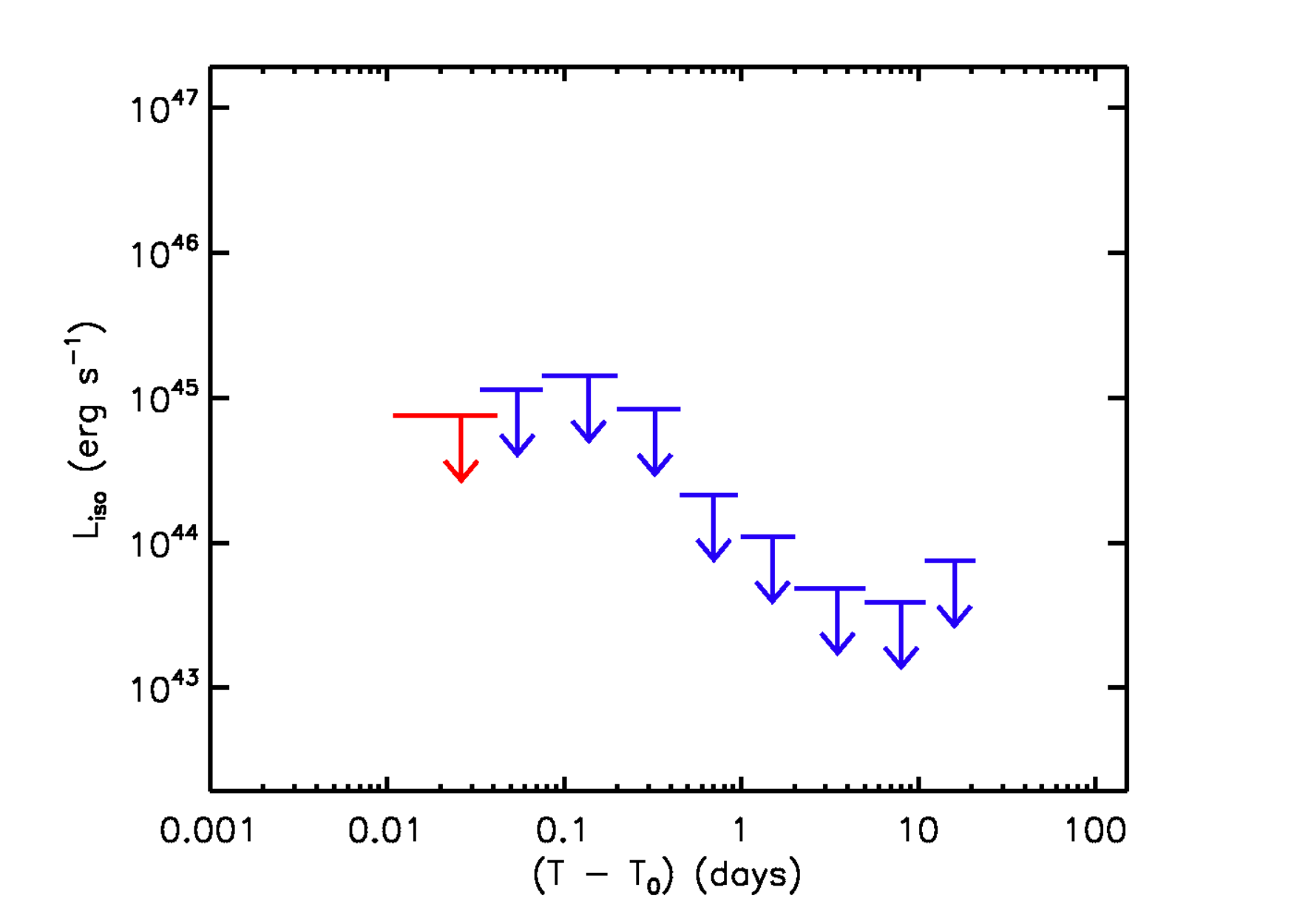}
}
   \caption{
The \agilep-GRID sequence of ULs converted into {\fvv isotropic} luminosity limits (in $\rm erg\, s^{-1}$) {\fvv assum}ing the OT distance of 40 Mpc{\fvv . Top panel: data regarding the intervals preceding the NS--NS coalescence. In red color is marked the UL obtained in "GRB detection mode" for an integration of 1915 s in the 30 MeV\,--\,10 GeV band. Bottom panel: flux ULs obtained for the intervals following the NS--NS coalescence. In red color the UL obtained in "GRB detection mode" for an integration of 2665 s in the 30 MeV\,--\,10 GeV band. In both panels in blue color we show the ULs obtained with the ML analysis in the range 100 MeV -- 10 GeV}. All GRID ULs were obtained at the {\fvv OT} position.}
 \label{fig-6}
\end{figure}
one of the earliest upper limits to hard X-ray emission in the
band 18--60 keV at $\Delta t = 0.53 \, \rm d$ (see also Table 4 of
MMA17), $F_X < 3 \cdot 10^{-9} \, \rm erg \, cm^{-2}\, s^{-1}$,
which translates into a limiting isotropic luminosity of $L_{iso,X}
= 5.8 \cdot 10^{44} \, \rm erg \, s^{-1}$. We note that both GRID
and SA provide similar limits to the emitted luminosity
in their different energy bands even though measured at different times
(+0.011 days and +0.53 days, respectively). The non-imaging MCAL
provided  data as the \gw LR was de-occulted by the Earth.
\begin{figure}  [h!t]
   \centerline{\includegraphics[width=9.5cm]{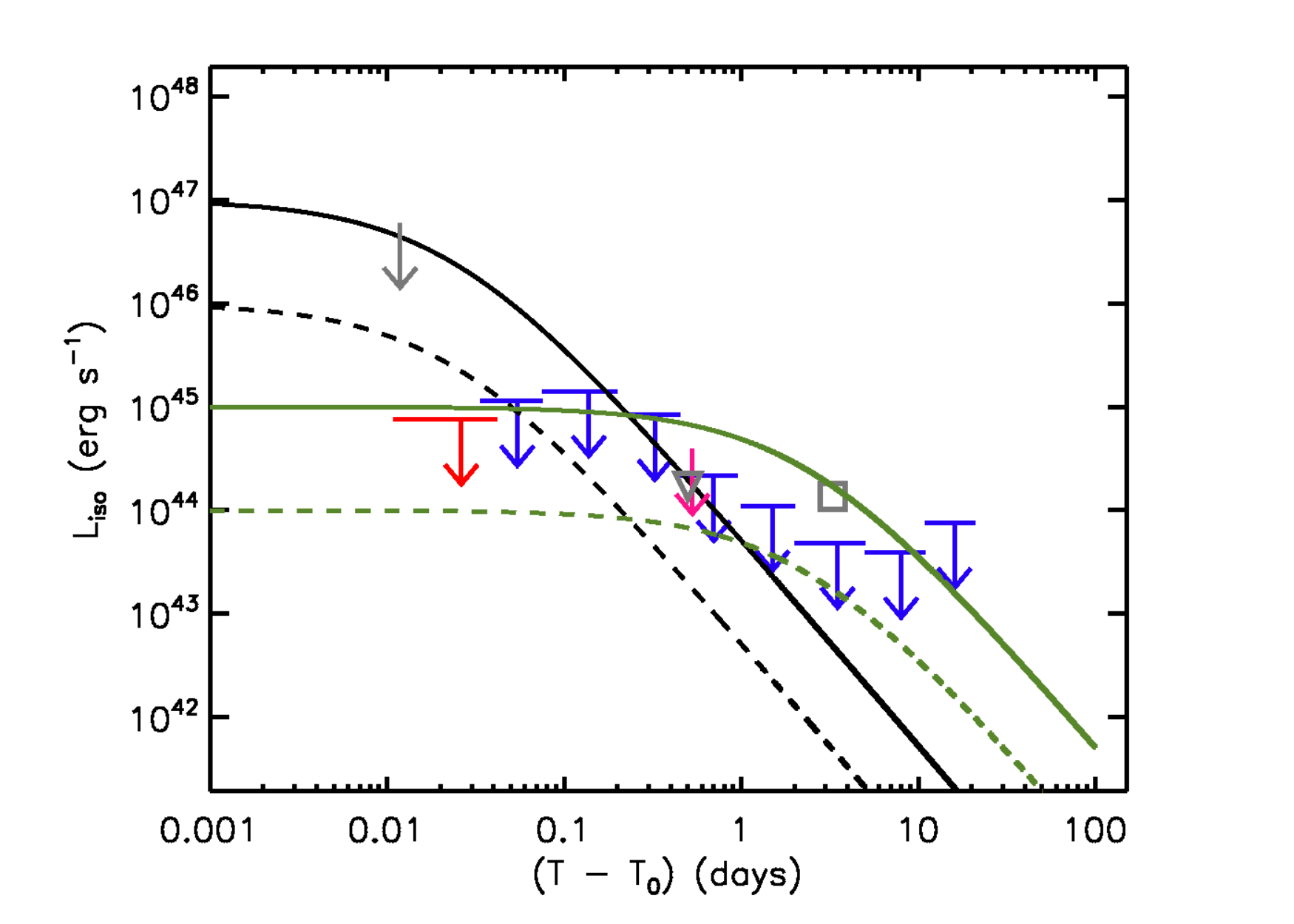}}
\caption{{\mm \agile 2\,$\sigma$ ULs obtained from observations of
the \gw localization region and converted into luminosity limits
(in $\rm erg\, s^{-1}$) for an OT distance of 40 Mpc}. In
blue color we show the \agilep-GRID ULs in the energy range 100 MeV\,
--\, 10 GeV. In red color we show the early GRID UL in the range 30
MeV\,--\, 10 GeV. 
 {\fv Marked} in {\fvv magenta} and gray colors are the SA and MCAL
flux ULs in the 18\,--\,60 $\rm keV$ and 400 $\rm keV$\,--\,100 $\rm
MeV$ bands, respectively. {\fvv We mark by a gray triangle 
the post-merger UL \fermigbm (20\,--\,100 $\rm keV$), and by a gray square the \intspi (500\,--\,1000 $\rm keV$), post-merger luminosity UL (MMA17).}
We also show the {\mm high-energy } luminosity curves relative to
the magnetar-like remnant model described in Section 4. The black
lines refer to a model with a poloidal magnetic field
$B_p$\,=\,$10^{15} \, \rm G$ and a radiation efficiencies $\xi$ of
$10^{-2}$ (solid line) and $10^{-3}$ (dashed line). Green lines
correspond to a model with B$_p$\,=\,$10^{14} \rm \, G$ and
$\xi\,=\,10^{-2}$ (solid) or $\xi\,=\,10^{-3}$ (dashed). }
\label{fig-7}
\end{figure}
Assuming the same spectral model applicable to GRB 170817A (MMA17),
the 90\% c.l. fluence upper limit is $F_{MCAL} < 3 \cdot 10^{-7}
\, \rm erg \, cm^{-2}$, which turns out to be similar to other
measurements by other calorimetric detectors (see Table 3 of MMA17).

{\mt The limiting luminosities implied by our measurements can be
interpreted in a context of high-energy radiation possibly emitted
by either an expanding fireball/jet or by a remnant left over from 
the NS--NS coalescence. The latter hypothesis is more interesting
in terms of constraints that can be obtained by the \agilep-GRID
observations. If the remnant is a magnetar-like system loaded with
a residual magnetic field and rapidly rotating \cite[e.g.][]
{1992Duncan,1992Usov,1994Thomp,1999Spruit,2001Zhang}, we can
constrain its magnetic field assuming initial millisecond spin
periods.} The electromagnetic emission (EM) by magnetic dipole
radiation of a star of radius $R$ with a poloidal magnetic field
$B_p$ and angular frequency $\Omega = 2 \, \pi/P$ (with $P$ as the
spin period) is $L(t) = B_p^2 \, R^6 \, \Omega(t)^4/6 \, c^3$
(with $c$ as the speed of light). Neglecting GW radiation at late
times after coalescence, integration of the energy loss equation
leads to the dependence of $\Omega$ as a function of time, $\Omega
(t) = \Omega_0 (1 + t/\tau)^{-1/2}$, with $\Omega_0$ as the initial
frequency and $\tau = 3 c ^3 I / ( B_p^2 R^6 \Omega_0^2) \simeq
(2 \cdot 10^3 {\rm \, s}) \, I_{45} \, B_{p,15}^{-2} \, P_{0,-3}^2
\, R_6^{-6}$, where $I_{45} $ is the compact object moment of
inertia  in units of $10^{45} \, {\rm g \, cm^2}$, $B_{p,15} =
B_p/10^{15} \, \rm G$, $P_{0,-3}$ is the initial spin period in
units of $10^{-3} \, \rm s$, and $R_6$ is the radius of the
compact object in units of $10^6$ cm.
For the EM-dominated regime of energy loss, we obtain the temporal
behavior of the spin-down luminosity \cite[e.g.][]{2001Zhang}
$L_{sd} (t) = L_0 / (1 + t/\tau)^2$, with $L_0 = I \,
\Omega^2_0 / (2 \, \tau) \simeq (10^{49} \, {\rm erg \, s^{-1}})
\, B_{p,15}^{2} \, P_{0,-3}^{-4} \, R_6^{6}$. It is interesting to
note that in the absence of absorption effects, the radiated
luminosity (in first approximation assumed here to be isotropic)
has the limiting behaviors $L_{sd} = L_0 $ for $t \ll \tau$ and
$L_{sd} = L_0 \, (t/\tau)^{-2}$ for $t \gg \tau$. In our case, we
can assume the remnant of radius $R_6 = 1$ and moment of inertia
$I_{45} = 1$ to rotate with an initial millisecond spin period,
$P_{0,-3} = 1$. We then have the critical time $\tau$ depending
only on the surface magnetic field $\tau \simeq  (2 \cdot 10^3
{\rm \, s}) \, B_{p,15}^{-2}$.

Assuming a conversion of spin-down luminosity $L_{sd} (t) $ into
radiation by a factor $\xi$,
we show the radiated luminosity $L_{rad} (t) = \xi \, L_{sd}(t) $
in Figure~\ref{fig-7} under the assumptions of two different values
of the poloidal magnetic field $B_p = 10^{15} \, \rm G$ and
$10^{14} \, \rm G$, for two values of the conversion factors, $\xi
= 10^{-2}$ and $\xi = 10^{-3}$. {\mm We denote by $\xi$ the
possibly different conversion efficiencies into hard X-rays and
the GRID energy range above 30 MeV.}
We show in
Figure~\ref{fig-7} all \agile upper limits (from MCAL, SA 
and GRID) together with the temporal behavior of high-energy
radiation expected from a rapidly rotating magnetar{\fvv , compared with 
the post-merger flaring UL from \fermigbm (20\,--\,100 keV) and \intspi (500\,--\,1000 $\rm keV$; MMA17) 
for integrations of 1 day and 4.7 days, respectively}.
 We see that the early  \agilep-GRID upper limits 
are important in general and, in particular, exclude highly magnetized
magnetars. We see in Figure~\ref{fig-7} that magnetar models with
$B_p = 10^{15} \, \rm G$ are excluded by the earliest GRID upper
limit with high confidence also for $\xi = 10^{-3}$. Analogously,
 models with $B_p = 10^{14} \, \rm G$ are excluded for $\xi =
10^{-2}$.

Constraining the gamma-ray emission of \gw at early times (within
the first few thousand seconds) is relevant also to models of
possible magnetic field reconnection following the formation of a
black hole. Transient gamma-ray emission can be expected from the
collapse of the coalescing NSs into a newly born black hole. We
exclude models envisioning gamma-ray luminosities near $10^{45} \,
\rm erg \, s^{-1}$ at +1000 s after coalescence.

\agile continues to operate nominally and will continue to observe
the high-energy sky in the search for transient events associated
with gravitational-wave events.

\vskip 0.3in \acknowledgements

{\mm This work is part of the multi-frequency observation campaign
of GW170817, whose main results are summarized in the
multi-messenger astronomy (MMA) paper. We warmly thank our
\lvc and MMA collaborators for sharing information on the
event and follow-up observations.{ \fvv We thank the anonymous referee for valuable comments.}}
 \agile is an ASI space mission developed with programmatic support by INAF and INFN.
We thank INAF and ASI for support.
 We acknowledge partial support through the ASI grant No. I/028/12/2.

\end{document}